\documentclass[conference,final]{IEEEtran}

\usepackage{acronym}
\usepackage{amsfonts}
\usepackage[dvips]{graphicx}
\usepackage{times}
\usepackage{cite}
\usepackage{amsmath}
\usepackage{array}
\usepackage{amssymb}
\usepackage{stfloats}
\usepackage{slashbox}
\usepackage{graphicx}
\usepackage{footnote}
\usepackage{amsthm}
\usepackage{booktabs}
\usepackage{array}
\usepackage{algorithmic}
\usepackage{algorithm}
\usepackage{subeqnarray}
\usepackage{cases}
\usepackage{threeparttable}
\usepackage{color}
\usepackage{epstopdf}
\usepackage{multirow}
\usepackage{tabularx}
\usepackage{enumerate}
\usepackage{multicol}
\usepackage{subcaption}

\newcommand{\figref}[1]{{Fig.}~\ref{#1}}

\def\bb0{{\mathbb{0}}}

\def\bb{{\mathbf{b}}}

\def\bh{{\mathbf{h}}}

\def\bv{{\mathbf{v}}}

\def\b0{{\mathbf{0}}}

\def\bH{{\mathbf{H}}}

\def\sf0{{\mathsf{0}}}

\def\rm0{{\mathrm{0}}}

\def\Nt{{N_\mathrm{t}}}

\def\Pt{{P_{\mr{t}}}}

\newcommand{\mb}{\mathbf}
\newcommand{\mr}{\mathrm}

\acrodef{CSI}[CSI]{channel state information}
\acrodef{CSIT}[CSIT]{channel state information at the transmitter}
\acrodef{CSIR}[CSIR]{channel state information at the receiver}
\acrodef{MIMO}[MIMO]{multiple-input multiple-output}
\acrodef{SISO}[SISO]{single-input single-output}
\acrodef{MISO}[MISO]{multiple-input single-output}
\acrodef{SIMO}[SIMO]{single-input multiple-output}
\acrodef{ADCs}[ADCs]{analog-to-digital convertors}
\acrodef{SNR}[SNR]{signal-to-noise ratio}
\acrodef{AWGN}[AWGN]{additive white Gaussian noise}
\acrodef{MRT}[MRT]{maximal ratio transmission}

\IEEEoverridecommandlockouts
\begin{document}
%

\title{Limited Feedback in MISO Systems with Finite-Bit ADCs}
\author{
	\IEEEauthorblockN{Jianhua~Mo and Robert W. Heath Jr.}
	\IEEEauthorblockA{Wireless Networking and Communications Group\\
	The University of Texas at Austin, Austin, TX 78712, USA}
	Email: \{jhmo, rheath\}@utexas.edu
	\thanks{This material is based upon work supported in part by the National Science Foundation under Grant No. NSF-CCF-1319556 and NSF-CCF-1527079.}
}
\maketitle
\begin{abstract}
We analyze limited feedback in systems where a multiple-antenna transmitter sends signals to single-antenna receivers with finite-bit ADCs. If channel state information (CSI) is not available with high resolution at the transmitter and the precoding is not well designed, the inter-user interference is a big decoding challenge for receivers with low-resolution quantization. In this paper, we derive achievable rates with finite-bit ADCs and finite-bit CSI feedback. The performance loss compared to the case with perfect CSI is then analyzed. The results show that the number of bits per feedback should increase linearly with the ADC resolution to restrict the loss.
\end{abstract}

\section{Introduction}

The wide bandwidth and large antenna arrays in future communication systems impose big challenges for the hardware design of the receiver, which has to efficiently process multiple signals from antennas at a much faster pace. The analog-to-digital converter (ADC) is one of the bottlenecks. At rates above 100 Mega samples per second, ADC power consumption increases quadratically with the sampling frequency \cite{Murmann_FTFC13, Murmann_16}.


The use of few- and especially one-bit ADCs is proposed as one approach to overcoming this challenge, for example, in the millimeter wave \ac{MIMO} channel \cite{Singh_TCOM09, Mezghani_ISIT07, Dabeer_ICC10, Mezghani_WSA10, Mezghani_ISIT12, Bai_Qing_ETT15, Mo_Jianhua_TSP15, Mo_Jianhua_arxiv16a, Mo_Jianhua_arxiv16b, Wen_Chao-Kai_TSP16} and massive \ac{MIMO} channel \cite{Choi_TCOM16, Mollen_TWC16, Studer_TCOM16, Jacobsson_ICC15, Fan_Li_CL15, Wang_Shengchu_TWC15}. This work has shown that low resolution ADCs can be used in practical communications. For instance, it is found that there is nearly no performance loss (less than $2$ dB) at low SNR compared to infinite-bit ADCs; it is possible to estimate the channel (IID Rayleigh fading or correlated) and detect symbols (QPSK or higher-order QAM) with coarse quantization.

In our previous work \cite{Mo_Jianhua_Asilomar15}, we analyzed the single-user \ac{SISO} and \ac{MISO} channels with \emph{one-bit} ADCs where the the transmitter sends the capacity-achieving QPSK symbols. Our proposed codebook design for the \ac{MISO} beamforming case separately quantizes the channel direction and the residual phase to incorporate the phase sensitivity of QPSK symbols. This design, however, cannot be extended to the channel with more than one-bit ADCs because the optimal signaling in this case is unknown.

In this paper, we assume that the transmitter adopts suboptimal Gaussian signaling. Since Gaussian signaling is circular symmetric, a single codebook quantizing the channel direction is enough. We derive the bounds of achievable rates with finite-bit ADC and finite-bit feedback. The rate and power losses incurred by the finite rate feedback compared to perfect CSIT is analyzed. Our results bridge the gap between the case of infinite-bit ADC \cite{Jindal_IT06} and one-bit ADC \cite{Mo_Jianhua_Asilomar15}.


\section{System Model}

\begin{figure}
    \begin{subfigure}{.49\textwidth}
        \centering
        \includegraphics[width=\linewidth]{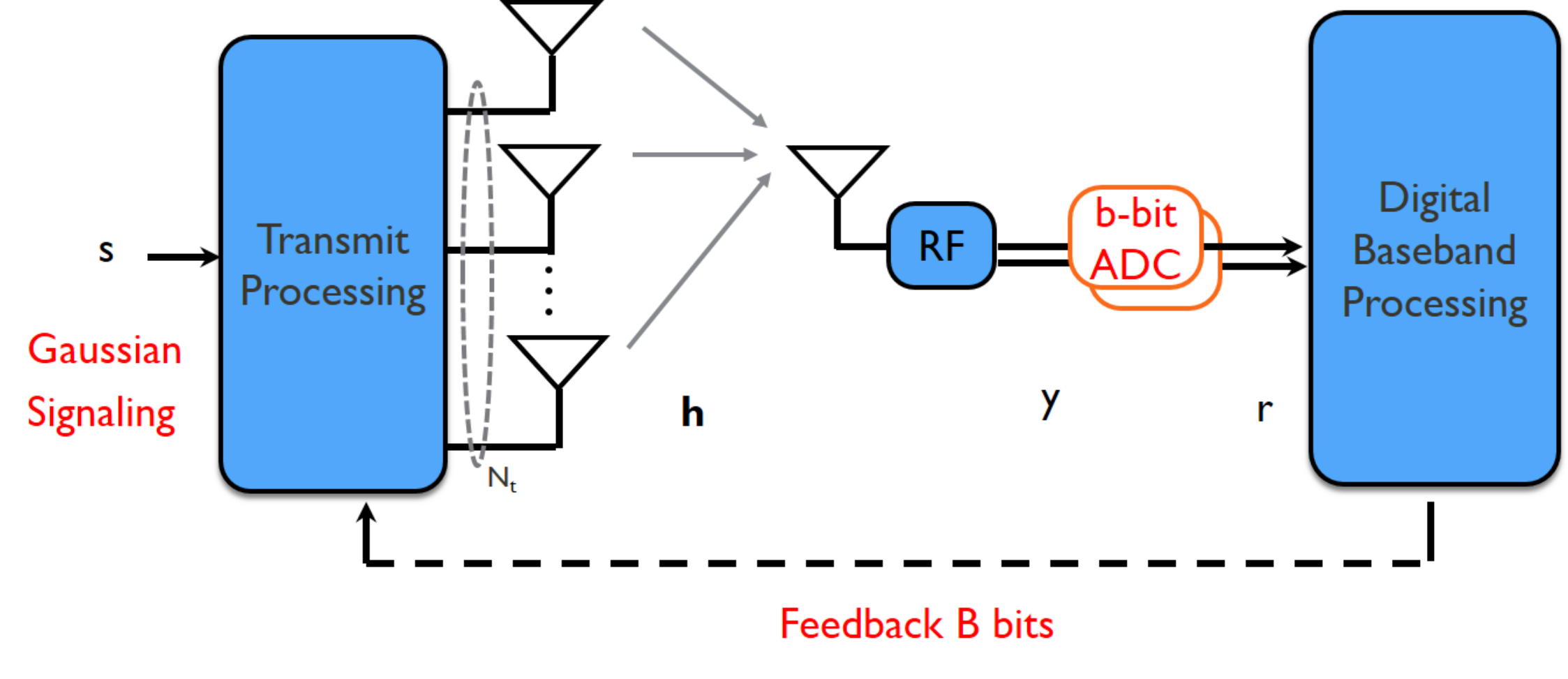}
        \caption{Single-user MISO system} \label{fig:SU_MISO}
    \end{subfigure}\hfill
    \begin{subfigure}{.49\textwidth}
        \centering
        \includegraphics[width=0.95\linewidth]{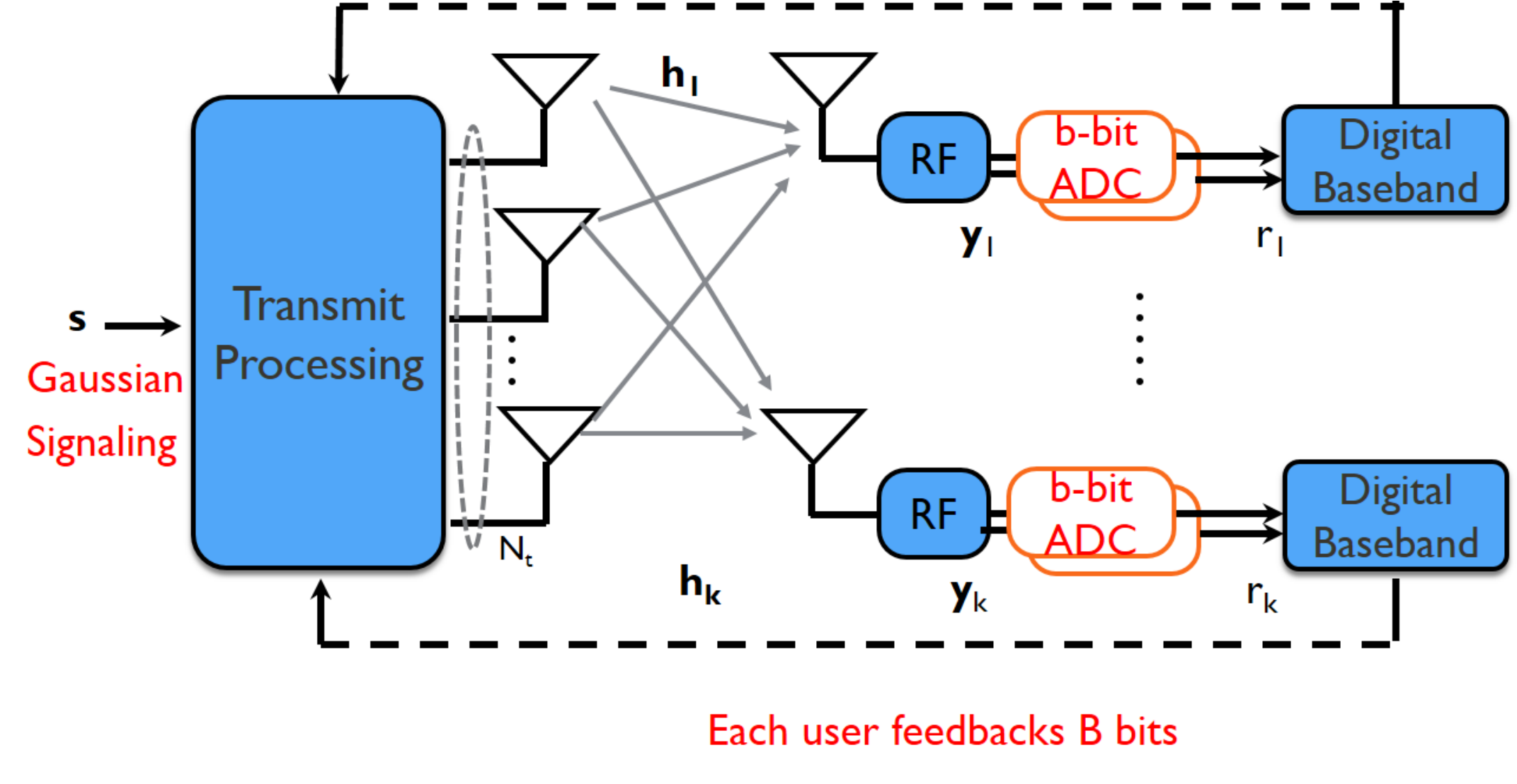}
        \caption{Multi-user MISO system} \label{fig:MU_MISO}
    \end{subfigure}
    \caption{MISO systems with finite-bit quantization and limited feedback. At each receiver, there are two $b$-bit ADCs. There is also a low-rate feedback path from each receiver to the transmitter.} \label{fig:MISO}
\end{figure}

In this paper, we consider single-user and multiple-user MISO systems shown in \figref{fig:MISO}. The transmitter is equipped with $\Nt$ antennas, while each receiver has only one antenna with finite-bit ADCs. In our system, there are two $b$-bit resolution quantizers that separately quantize the real and imaginary part of the received signal. We assume that uniform quantization is applied since it is easier for implementation and achieves only slightly worse performance than non-uniform case\cite{Max_IRE60}.

We assume that $B$  bits are used to convey the channel direction information. A codebook $\mathcal{C} = \left\{\widehat{\bh}^{(0)}, \widehat{\bh}^{(1)}, \cdots, \widehat{\bh}^{(2^{B}-1)} \right\}$ is shared by the transmitter and receiver. The receiver sends back the index $i$ of $\widehat{\bh}^{(i)}$ maximizing $|\bh^* \widehat{\bh}^{(i)}|$ where $\bh$ represents the channel. Then the transmitter performs beamforming based on the feedback information. Similar to a MISO system with infinite-resolution ADCs, random vector quantization (RVQ), which performs close to the optimal quantization and is amenable to analysis \cite{Jindal_IT06, Au-Yeung_TWC07}, is adopted to quantize the direction of channel $\bh$. In the codebook $\mathcal{C}$, each of the quantization vectors is independently chosen from the isotropic distribution on the Grassmannian manifold $\mathcal{G}(\Nt, 1)$ \cite{Love_IT03}.

Different from our previous work \cite{Mo_Jianhua_Asilomar15} where capacity-achieving QPSK signaling was adopted, we assume that Gaussian signaling is used at the transmitter. Although Gaussian signalling is suboptimal, it is amenable for analyses and close to optimal at low and medium SNR \cite{Mezghani_ISIT12, Mo_Jianhua_TSP15}.

In this paper, we assume the channel follows IID Rayleigh fading. The extension to correlated channel model is an interesting topic for future work.
We also assume the receiver has perfect channel state information. This is justified by the prior work on channel estimation with low resolution ADCs, for example \cite{Dabeer_ICC10, Mezghani_WSA10, Wen_Chao-Kai_TSP16, Mo_Jianhua_arxiv16b}. Furthermore, the feedback is assumed to be delay and error free, as is typical in limited feedback problems. 

\section{Single-user MISO Channel with Finite-bit ADCs and Limited Feedback} \label{sec:SU_MISO}

We first consider a single-user MISO system with finite-bit quantization, as shown in Fig. \ref{fig:SU_MISO}.
Assuming perfect synchronization and a narrowband channel, the baseband received signal in this MISO system is
\begin{equation}
y = \mb{h}^*\mb{v} s + n,
\end{equation}
where $\mb{h} \in \mathbb{C}^{\Nt\times 1}$ is the channel vector, $\mb{v}\in \mathbb{C}^{\Nt \times 1} (\| \bv\|=1)$ is the beamforming vector, $s$ is the Gaussian distributed symbol sent by the transmitter, $y\in \mathbb{C}$ is the received signal before quantization, and $n \sim \mathcal{CN}(0, \sigma^2)$ is the circularly symmetric complex Gaussian noise. The average transmit power is $\Pt$, i.e., $\mathbb{E}[|s|^2] = \Pt$.

The output after the finite-bit quantization is
\begin{equation}
r = \mathcal{Q}\left(y\right) = \mathcal{Q} \left(\mb{h}^*\mb{v} s + n\right),
\end{equation}
where $\mathcal{Q}(\cdot)$ is the finite-bit quantization function applied separately to the real and imaginary parts.

\begin{table*}
	\centering
	\caption{The optimum uniform quantizer for a Gaussian unit-variance input signal \cite{Max_IRE60} }
	\label{tab:Eta_b}
	\begin{tabular}{|c|c|c|c|c|c|c|c|c|}
		\hline
		Resolution $b$  & 1-bit  & 2-bit  & 3-bit & 4-bit & 5-bit & 6-bit & 7-bit & 8-bit\\
		\hline
		NMSE $\eta_b$ & $\frac{\pi-2}{\pi} \left(\approx 0.3634 \right)$ & 0.1175 & 0.03454 & 0.009497 & 0.002499 & 0.0006642 & 0.0001660 & 0.00004151\\
		\hline
        $10 \log_{10} \left(1 -\eta_b \right)$ & -1.9613  & -0.5429  & -0.1527  & -0.0414  & -0.0109  & -0.0029  & -0.0007 &  -0.0002 \\
        \hline
        $\log_2 \left(\frac{1}{\eta_b}\right) $ & 1.46 & 3.09 &  4.86 &  6.72 &  8.64 &  10.56  & 12.56 &  14.56\\
        \hline
	\end{tabular}
\end{table*}

%

By Bussgang's theorem \cite{Bussgang_52, Fletcher_JSTSP07, Mezghani_ISIT12}, the quantization output can be decoupled into two uncorrelated parts, i.e.,
\begin{IEEEeqnarray}{rCl}
r &=& (1-\eta_b) y + n_{\mr{Q}} \\
& =& (1-\eta_b) \bh^* \bv s + (1-\eta_b) n + n_{\mr{Q}},
\end{IEEEeqnarray}
where $\eta_b = \frac{\mathbb{E}[|r-y|^2]}{\mathbb{E}[|y|^2]}$ is the normalized mean squared error and $n_{\mathrm{Q}}$ is the quantization noise with variance $\sigma_{\mr{Q}}^2 = \eta_b (1-\eta_b) \mathbb{E}[|y|^2] = \eta_b(1-\eta_b) \left( | \bh^* \bv |^2 \Pt + \sigma^2 \right)$. Therefore, the effective noise $n_{\mathrm{ef}} \triangleq (1-\eta_b) n + n_{\mr{Q}}$ has variance $\eta_b(1-\eta_b) \left( | \bh^* \bv |^2 \Pt \right) + (1-\eta_b) \sigma^2 $. The values of $\eta_b \left(1\leq b \leq 8\right)$ are listed in Table \ref{tab:Eta_b}.
The resulting signal-to-quantization and noise ratio (SQNR) at the receiver is
\begin{IEEEeqnarray*}{rCl}
\mr{SQNR}
&=& \frac{ \left(1-\eta_b \right)^2 \Pt \left|\bh^*\bv \right|^2 }{ \left( 1-\eta_b \right)^2 \sigma^2 + \sigma_{\mr{Q}}^2 } = \frac{ \left(1-\eta_b \right) \Pt \left|\bh^*\bv \right|^2 }{ \eta_b  \Pt \left|\bh^*\bv \right|^2 + \sigma^2 }. \IEEEyesnumber
\end{IEEEeqnarray*}

Denote the achievable rate with $b$-bit ADC and $B$-bit feedback as $R(b,B)$. In this paper, `$b=\infty$' represents the case of full-precision ADCs, while `$B=\infty$' represents the case of perfect CSIT. Assuming that the noise $n_{\mathrm{Q}}$ follows the worst-case Gaussian distribution, the average achievable rate with perfect CSIT and conjugate beamforming is

\begin{IEEEeqnarray}{rCl}
R(b, \infty) &=& \mathbb{E}_{\bh} \left[\log_2 \left(1+ \frac{ \left(1-\eta_b \right) \Pt \left\|\bh\right\|^2 }{ \eta_b \Pt \left\|\bh \right\|^2 + \sigma^2} \right) \right] \\
&\stackrel{(a)}{\leq} & \log_2 \left(1+ \frac{ \left(1-\eta_b \right) \Pt \mathbb{E} \left[\left\|\bh\right\|^2 \right] }{ \eta_b \Pt \mathbb{E} \left[ \left\|\bh \right\|^2 \right] + \sigma^2} \right) \\
& \stackrel{(b)}{=} & \log_2 \left(1+ \frac{ \left(1-\eta_b \right) \Pt \Nt }{ \eta_b \Pt \Nt + \sigma^2} \right)
\end{IEEEeqnarray}
where $(a)$ follows from the concavity of the function $f(x)=\log_2 \left(1 + \frac{ax}{bx+c}\right) (a>0, b>0, c>0)$ when $x>0$, $(b)$ follows from the assumption of IID Rayleigh fading channel.

In the low and high SNR $\left(\frac{\Pt}{\sigma^2}\right)$ regimes, the average achievable rate with perfect CSIT is approximated as,
\begin{align} \label{eq:SU_Rate_CSIT_low_high_SNR}
	R(b, \infty) & \approx \left\{
	\begin{array}{lc}
	\log_2 \left(1+ \frac{ \left(1-\eta_b \right) \Pt \Nt }{  \sigma^2} \right), &\text{when $\frac{\Pt}{\sigma^2}$ is small,} \\
	\log_2 \left( \frac{1}{\eta_b}\right), &\text{when $\frac{\Pt}{\sigma^2}$ is large.}
	\end{array}
	\right.
\end{align}

It is seen that the high SNR rate is limited by the signal-to-quantization ratio (SQR) defined as $\mr{SQR} \triangleq \frac{1}{\eta_b}$.
Since $\eta_b \approx \frac{\pi \sqrt{3}}{2} 2^{-2b}$ when $b\geq 3$ \cite{Gersho_Book12}, the achievable rate at high SNR is
\begin{IEEEeqnarray}{rCl}
	R(b, \infty)
	&\approx & 2 b - \log_2 \frac{\pi \sqrt{3}} {2} \\
	&\approx & 2  b - 1.44 \quad \text{bps/Hz}.
\end{IEEEeqnarray}
The values of $\log_2 \left( \frac{1}{\eta_b}\right)$ are given in Table \ref{tab:Eta_b}.

Averaging over the RVQ codebooks, the achievable rate under limited feedback is
\begin{IEEEeqnarray}{rCl}
& &R(b, B) \\
&=& \mathbb{E}_{\bh, \mathcal{C}} \left[\log_2 \left(1+ \frac{ \left(1-\eta_b \right) \Pt \left|\bh^*\bv \right|^2 }{  \eta_b \Pt \left|\bh^*\bv \right|^2+\sigma^2} \right) \right] \\
&\stackrel{(a)}{\approx}  & \log_2 \left(1+ \frac{ \left(1-\eta_b \right) \Pt \Nt \left(1- 2^B \beta\left(2^B, \frac{\Nt}{\Nt-1}\right) \right) }{  \eta_b \Pt \Nt \left(1- 2^B \beta\left(2^B, \frac{\Nt}{\Nt-1}\right) \right)+\sigma^2} \right) \IEEEeqnarraynumspace \\
&\stackrel{(b)}{\geq} & \log_2 \left(1+ \frac{ \left(1-\eta_b \right) \Pt \Nt \left(1- 2^{-\frac{B}{\Nt-1}} \right) }{  \eta_b \Pt \Nt \left(1- 2^{-\frac{B}{\Nt-1}} \right)+\sigma^2} \right)
\end{IEEEeqnarray}
where $\beta(\cdot, \cdot)$ is a beta function. $(a)$ follows from $\mathbb{E}\left[\|\bh \|^2 \right] = \Nt$ and $\cos^2 \left(\angle \left(\bh, \bv\right) \right) = 1 - 2^B \beta\left(2^B, \frac{\Nt}{\Nt-1} \right)$ \cite{Au-Yeung_TWC07},
$(b)$ follows from the inequality $2^B \beta\left(2^B, \frac{\Nt}{\Nt-1}\right) \leq 2^{- \frac{B}{\Nt-1}}$ \cite{Jindal_IT06}.

In the low and high SNR regimes, the average achievable rate with limited feedback is
\begin{align} \label{eq:SU_Rate_fb_low_high_SNR}
   & R(b, B) \notag \\
   &\approx\left\{
	\begin{array}{lc}
	\log_2 \left(1+ \frac{ \left(1-\eta_b \right) \Pt \Nt \left( 1- 2^{-\frac{B}{\Nt-1}} \right) }{  \sigma^2}  \right), &\text{if $\frac{\Pt}{\sigma^2}$ is small,} \\ 
	\log_2 \left( \frac{1}{\eta_b}\right), & \text{if $\frac{\Pt}{\sigma^2}$ is large.}  
	\end{array}
	\right.
\end{align}

%
%

Comparing $R(b, \infty)$ in \eqref{eq:SU_Rate_CSIT_low_high_SNR} and $R(b, B)$ in \eqref{eq:SU_Rate_fb_low_high_SNR}, we find that at low SNR, the power loss between $R(b, \infty)$ and $R(b, B)$ is $\approx 10 \log_{10} \left( 1- 2^{-\frac{B}{\Nt-1}} \right)$ dB. The result is similar to the case with infinite-bit ADCs \cite{Au-Yeung_TWC07, Mukkavilli_IT03}.
In contrary, at high SNR, both $R(b, \infty)$ and $R(b, B)$ approach the same upper bound and the rate loss due to limited feedback is zero.

At last, the achievable rate with infinite-bit ADC and perfect CSIT is known as $R(\infty, \infty)= \log_2 \left(1+ \frac{\Pt \Nt}{\sigma^2} \right)$. We find that at low SNR, the power loss incurred by the finite-bit ADC is $10 \log_{10} \left(1 -\eta_b\right)$ dB while that by limited feedback is $10 \log_{10} \left(1 - 2^{-\frac{B}{\Nt-1}}\right)$ dB.

\section{Multi-User MISO Channel with Finite-bit ADCs and Limited Feedback}\label{sec:MU_MISO}

We now consider a multi-user MISO channel shown in \figref{fig:MU_MISO} where a $\Nt$-antenna transmitter sends signals to $K \left(1<K \leq \Nt\right)$ single-antenna receivers.
The quantization output at the $k$-th receiver is
\begin{IEEEeqnarray}{rCl}
r_k & =& \mathcal{Q} \left( \sqrt{\rho} \sum_{i=1}^{K}\bh_k^*\bv_i s_i  + n_k \right) \\
& =&  (1-\eta_{b}) \sqrt{\rho} \bh_k^*\bv_i s_i + (1-\eta_{b}) \sqrt{\rho} \sum_{i=1, i\neq k}^{K}\bh_k^*\bv_i s_i \nonumber \\
& &\negmedspace {} + (1-\eta_{b}) n_k +n_{\mr{Q}}
\end{IEEEeqnarray}
where $\rho \triangleq  \frac{\Pt}{K}$ is the power allocated to each user, $\bv_i$ is the beamforming vector for user $i$, $n_k \sim \mathcal{CN}(0, \sigma^2)$ is the circularly symmetric complex Gaussian noise, and the quantization noise $n_{\mr{Q}}$ has variance $\eta_{b} (1-\eta_{b})\left( \rho \sum_{i=1, i\neq k}^K \left|\bh_k^*\bv_i \right|^2 +\sigma^2\right)$. Therefore, the signal-to-interference, quantization and noise ratio (SIQNR) at the $k$-th receiver is
\begin{IEEEeqnarray*}{rCl}
\mr{SIQNR}_k
&=& \frac{ \left(1-\eta_b \right) \rho \left|\bh_k^*\bv_i \right|^2 }{  \eta_b \rho \left|\bh_k^*\bv_k \right|^2 + \rho \sum_{i=1, i\neq k}^K \left|\bh_k^*\bv_i \right|^2 +\sigma^2}. \IEEEyesnumber
\end{IEEEeqnarray*}


%
%
%

If there is perfect CSIT and the transmitter designs zero-forcing beamforming $\bv_i^{\mr{ZF}} \left(1\leq i \leq K\right)$ such that $\bh_k^* \bv_i^{\mr{ZF}}=0$ for $k\neq i$, the average rate per user is
\begin{IEEEeqnarray}{rCl}
	R^{\mr{ZF}}(b, \infty)
	&=& \mathbb{E}_{\bH} \left[\log_2 \left(1+ \frac{ \left(1-\eta_b \right) \rho \left|\bh_k^*\bv_k^{\mr{ZF}} \right|^2 }{ \eta_b \rho \left|\bh_k^*\bv_k^{\mr{ZF}} \right|^2 + \sigma^2} \right) \right] \\
    &\stackrel{(a)}{\leq}& \log_2 \left(1+ \frac{ \left(1-\eta_b \right) \rho (\Nt-K+1) }{ \eta_b \rho (\Nt-K+1) + \sigma^2} \right)
\end{IEEEeqnarray}
where $\mathbb{E}_{\bH}\left[\left|\bh_k^* \bv_k^{\mr{ZF}}\right|^2\right] = \mathbb{E}_{\bH}\left[\|\bh_k\|^2\right] \mathbb{E}_{\bH}\left[ |\widetilde{\bh}_k^*  \bv_k^{\mr{ZF}} |^2 \right] = \mathbb{E}_{\bH} \left[\|\bh_k\|^2\right] \mathbb{E}_{\bH} \left[ \cos ^2 \angle \left(\widetilde{\bh}_k, \bv_k^{\mathrm{ZF}}\right)\right]
 =\Nt \frac{\Nt-K+1}{\Nt} = \Nt-K+1$ where $\widetilde{\bh}_k = \frac{\bh}{\|\bh\|}$.

In the case without perfect CSIT, each receiver feeds back $B$ bits as the index of the quantized channel $\widehat{\bh}_k$, then the transmitter designs zero-forcing precoding based on $\widehat{\bh}_k \left(1\leq k \leq K\right)$. The average achievable rate is shown in \eqref{eq:MU_Rate_fb}-\eqref{eq:MU_Rate_fb_LB}.
\begin{figure*}[t]
\begin{IEEEeqnarray}{rCl}
R^{\mr{ZF}}\left(b, B\right)
& = & \mathbb{E}_{\bH, \mathcal{C}} \left[\log_2 \left(1+ \frac{ \left(1-\eta_b \right) \rho \left|\bh_k^*\bv_k \right|^2 }{  \eta_b \rho \left|\bh_k^*\bv_k \right|^2 + \rho \sum_{i=1, i\neq k}^K \left|\bh_k^*\bv_i \right|^2 +\sigma^2} \right) \right] \label{eq:MU_Rate_fb} \\
& \approx & \log_2 \left(1+ \frac{ \left(1-\eta_b \right) \rho \mathbb{E} \left[ \left|\bh_k^*\bv_k \right|^2 \right] }{  \eta_b \rho \mathbb{E} \left[ \left|\bh_k^*\bv_k \right|^2 \right] + \rho \sum_{i=1, i\neq k}^K \mathbb{E} \left[ \left|\bh_k^*\bv_i \right|^2 \right] +\sigma^2} \right) \\
& = & \log_2 \left(1+ \frac{ \left(1-\eta_b \right) \rho \left(\Nt-K+1\right) \left( 1-2^B \beta\left(2^B, \frac{\Nt}{\Nt-1}\right) \right) }{  \eta_b \rho \left(\Nt-K+1\right) \left( 1-2^B \beta\left(2^B, \frac{\Nt}{\Nt-1}\right) \right) + \rho \left(K-1\right)  \frac{\Nt}{\Nt-1} 2^B \beta\left(2^B, \frac{\Nt}{\Nt-1}\right) +\sigma^2} \right) \\
& \geq & \log_2 \left(1+ \frac{ \left(1-\eta_b \right) \rho \left(\Nt-K+1\right) \left( 1-2^{-\frac{B}{\Nt-1}} \right) }{  \eta_b \rho \left(\Nt-K+1\right) \left( 1-2^{-\frac{B}{\Nt-1}} \right) + \rho \left(K-1\right)  \frac{\Nt}{\Nt-1} 2^{-\frac{B}{\Nt-1}} +\sigma^2} \right) \label{eq:MU_Rate_fb_LB}
\end{IEEEeqnarray}
\hrulefill
\end{figure*}


In \eqref{eq:MU_Rate_fb}-\eqref{eq:MU_Rate_fb_LB}, we use the equality $\mathbb{E}_{\bH, \mathcal{C}} \left[ \left|\bh_k^*\bv_i \right|^2 \right] = \mathbb{E}_{\bH} \left[ \left\|\bh_k \right\|^2 \right] \mathbb{E}_{\bH, \mathcal{C}} \left[ \left|\widetilde{\bh}_k^*\bv_i \right|^2 \right] = \frac{\Nt}{\Nt-1} 2^B \beta\left(2^B, \frac{\Nt}{\Nt-1}\right)$ \cite{Jindal_IT06} and the lower bound of $\mathbb{E}_{\bH, \mathcal{C}} \left[ \left| \bh_k^* \bv_k \right|^2 \right]$ as follows.
\begin{IEEEeqnarray*}{rCl}
\mathbb{E} \left[ \left| \bh_k^* \bv_k \right|^2 \right]
&\geq & \mathbb{E} \left[ \left| \bh_k^* \widehat{\bh}_k \right|^2 \right] \mathbb{E} \left[ \left| \widehat{\bh}_k^* \bv_k \right|^2 \right] \\
&=& \mathbb{E} \left[ \left| \bh_k \right|^2 \right] \mathbb{E} \left[ \left| \widetilde{\bh}_k^* \widehat{\bh}_k \right|^2 \right] \mathbb{E} \left[ \left| \widehat{\bh}_k^* \bv_k \right|^2 \right] \IEEEyesnumber \\
&\stackrel{(a)}{=}& \Nt \left( 1-2^B \beta\left(2^B, \frac{\Nt}{\Nt-1}\right) \right) \frac{\Nt-K+1}{\Nt},
\end{IEEEeqnarray*}
where $(a)$ follows from the equalities $\mathbb{E} \left[ \left| \widetilde{\bh}_k^* \widehat{\bh}_k \right|^2 \right] = 1-2^B \beta\left(2^B, \frac{\Nt}{\Nt-1}\right)$ \cite{Au-Yeung_TWC07} and $\mathbb{E} \left[ \left| \widehat{\bh}_k^* \bv_k \right|^2 \right] = \frac{\Nt - K+1}{\Nt}$.

Therefore, the rate loss incurred by limited feedback is
\begin{IEEEeqnarray}{rCl}
    \Delta R^{\mathrm{ZF}}(b) & = & R^{\mr{ZF}}\left(b, \infty \right) - R^{\mr{ZF}}\left(b, B \right)
\end{IEEEeqnarray}
and has an upper bounded shown in \eqref{eq:MU_Rate_loss_ub}.
\begin{figure*}[t]
\begin{align} \label{eq:MU_Rate_loss_ub}
	\Delta \overline{R}^{\mathrm{ZF}}(b) & =  \log_2 \left(1+ \frac{ \left(1-\eta_b \right) \rho (\Nt-K+1) }{ \eta_b \rho (\Nt-K+1) + \sigma^2} \right) - \log_2 \left(1+ \frac{ \left(1-\eta_b \right) \rho \left(\Nt-K+1\right) \left( 1-2^{-\frac{B}{\Nt-1}} \right) }{  \eta_b \rho \left(\Nt-K+1\right) \left( 1-2^{-\frac{B}{\Nt-1}} \right) + \rho \frac{\left(K-1\right) \Nt}{\Nt-1} 2^{-\frac{B}{\Nt-1}} +\sigma^2} \right)
\end{align}
\hrule
\end{figure*}

When the SNR $\left(\frac{\Pt}{\sigma^2} =
\frac{K \rho}{\sigma^2}\right)$ is low, the performance loss is
\begin{IEEEeqnarray*}{rCl} \label{MU_MISO_Rate_loss_low_SNR}
    & & \Delta \overline{R}^{\mathrm{ZF}} (b) \\
    & \approx &  \log_2 \left(1+ \frac{ \left(1-\eta_b \right) \rho (\Nt-K+1) }{  \sigma^2} \right) \IEEEyesnumber \\
    & & - \log_2 \left(1+ \frac{ \left(1-\eta_b \right) \rho \left(\Nt-K+1\right) \left( 1-2^{-\frac{B}{\Nt-1}} \right) }{\sigma^2} \right).
\end{IEEEeqnarray*}
It is found there is a power loss $\approx 10 \log_{10} \left( 1-2^{-\frac{B}{\Nt-1}} \right)$ dB which is similar to the single-user case shown in Section \ref{sec:SU_MISO}.

At high SNR, the rate loss is
\begin{IEEEeqnarray}{rCl}
	\Delta \overline{R}^{\mr{ZF}}(b) & \approx & \log_2 \left( 1 + \frac{1 - \eta_b}{\eta_b } \frac{1}{\frac{C_1}{C_2} + 1}\right)
\end{IEEEeqnarray}
where $C_1 \triangleq \left(\Nt-K+1\right) \left( 1-2^{-\frac{B}{\Nt-1}} \right)$  and $C_2 \triangleq \frac{\left(K-1\right) \Nt}{\Nt-1} 2^{-\frac{B}{\Nt-1}}$. To guarantee that the rate loss is less than $D$, the number of feedback bits $B$ should be large enough such that $\frac{1 - \eta_b}{\eta_b } \frac{1}{C_1/C_2+1}<2^D-1$.

When $b \geq 3$, $1-\eta_b \approx 0$ as shown in Table \ref{tab:Eta_b}. If $B \gg \Nt-1$, $\frac{C_1}{C_2} +1 \approx \frac{(\Nt-K+1)(\Nt-1)}{\Nt (K-1)} 2^{\frac{B}{\Nt-1}}$. In this case, to keep the rate loss constant, we want the following term
\begin{align} \label{eq:MU_MISO_scaling_law}
	 \frac{1}{\eta_b 2^{\frac{B}{\Nt-1}}}  \approx  \frac{2}{\pi \sqrt{3}} 2^{2b} 2^{-\frac{B}{\Nt-1}}
	=  \frac{2}{\pi \sqrt{3}} 2^{2 \left( b - \frac{B}{2 \left( \Nt-1 \right)} \right)}
\end{align}
to be less than a constant.
Therefore, if the ADC resolution $b$ increase $1$ bit, the number of feedback bits $B$ should increase $2(\Nt-1)$.

\section{Simulation Results}


\begin{figure}[t]
\begin{centering}
\includegraphics[width=1\columnwidth]{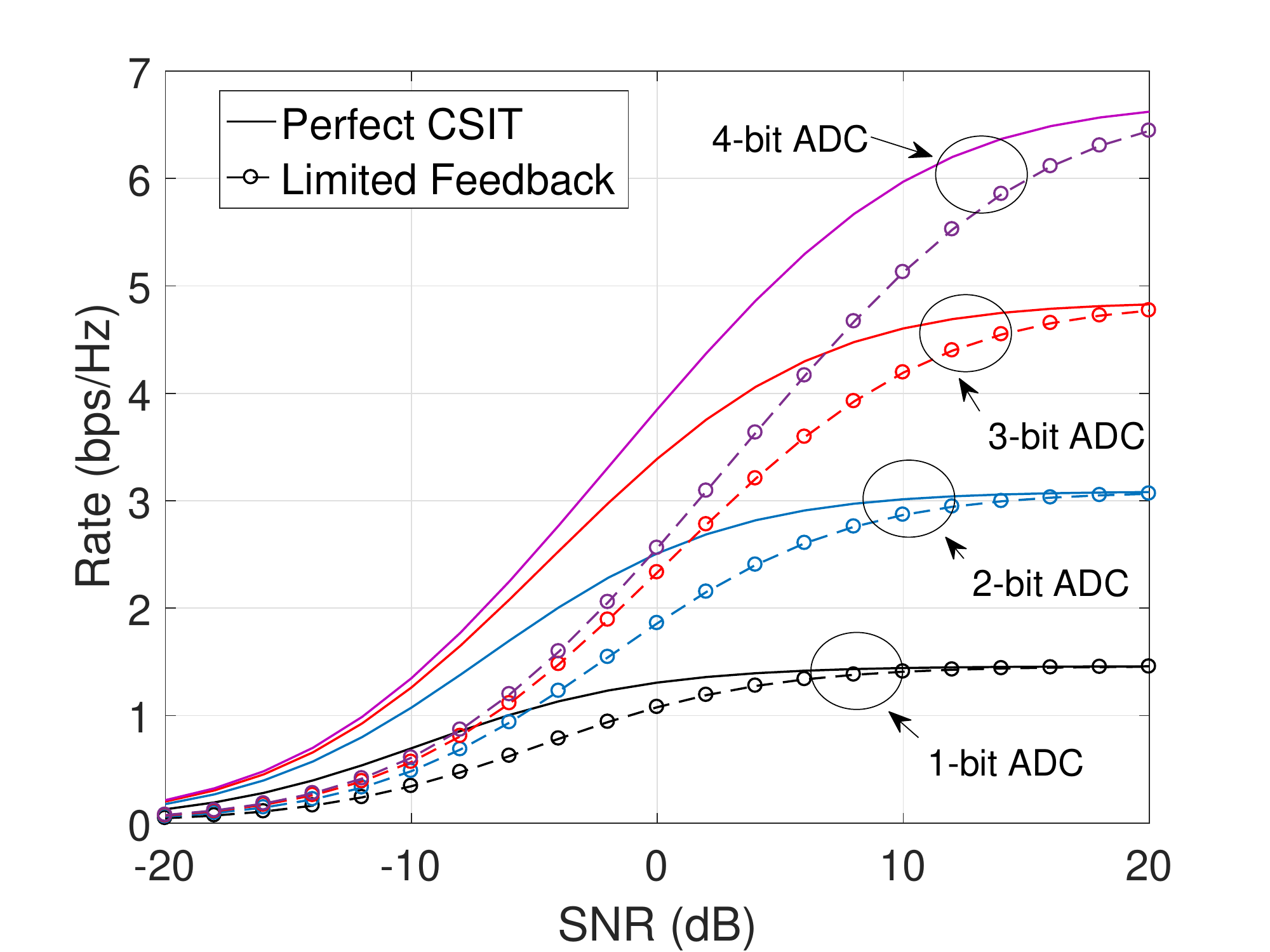}
\vspace{-0.1cm}
\centering
 \caption{The achievable rate of a single-user MISO system with CSIT and limited feedback when $\Nt=16$ and $B=8$.
 }\label{fig:SU_MISO_Rate_vs_SNR_Nt_16_B_8}
\end{centering}
\vspace{-0.3cm}
\end{figure}

\begin{figure}[t]
\begin{centering}
\includegraphics[width=1\columnwidth]{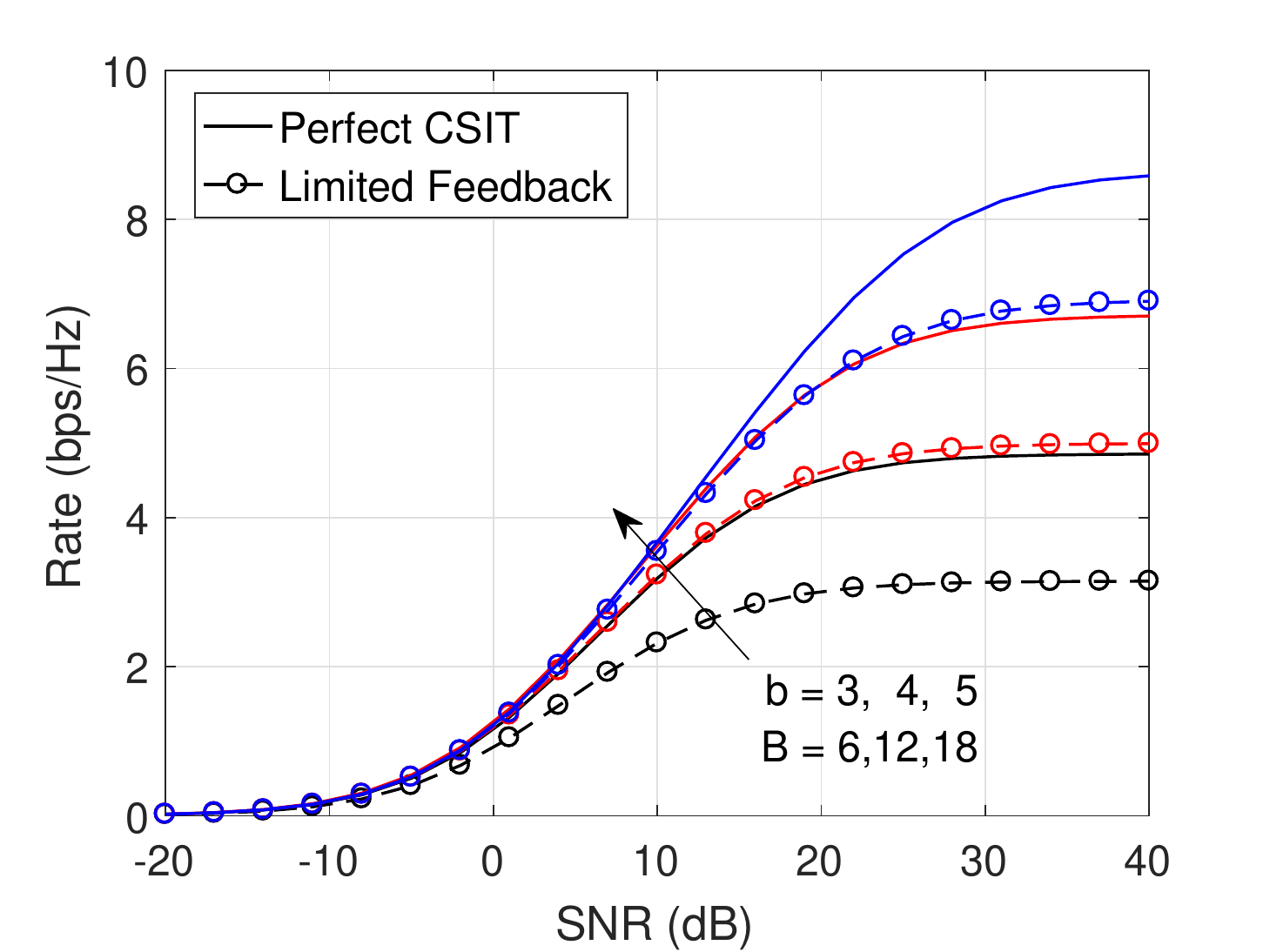}
\vspace{-0.1cm}
\centering
 \caption{The achievable rate of a multi-user MISO system with perfect CSIT and limited feedback when $\Nt=4$ and $K=2$. When there is perfect CSIT, the figure shows three cases where $b=3, 4, 5$. When there is limited feedback, the figure shows three cases where `$b=3, B=6$', `$b=4, B=12$' and `$b=5, B=18$'.
 }\label{fig:MU_MISO_Rate_vs_SNR_Nt_4_K_2}
\end{centering}
\vspace{-0.3cm}
\end{figure}

In this section, we compute the achievable rate for each channel realization then average over 1000 channel realizations with Rayleigh fading, i.e., $\bh_k \sim \mathcal{CN}(\mathbf{0}, \mathbf{I}_{\Nt})$. In the figures, $\mathrm{SNR (dB)} \triangleq 10 \log_{10} \frac{\Pt}{\sigma^2}$.

In \figref{fig:SU_MISO_Rate_vs_SNR_Nt_16_B_8}, we show the average achievable rates with perfect CSIT and limited feedback in a single-user MISO channel. First, the rate with perfect CSIT converges to $\log_2 \left( \frac{1}{\eta_b}\right)$, which is $1.46$, $3.09$, $4.86$, $6.72$ bps/Hz when $b=1, 2, 3, 4$. Note that these values are less than the theoretical upper bound $2b$ bps/Hz because Gaussian signaling is suboptimal.
Second, at high SNR (for instance, 10 dB when $b=1$, 20 dB when $b=4$), there is almost no rate loss between the perfect CSIT and limited feedback cases since the quantization noise dominates the AWGN noise in this regime. Third, in the low SNR regime ($<0$ dB), we see there is a constant horizontal distance between each pair of solid curve and dashed curve which implies that there is a constant power loss incurred by limited feedback. This is because the AWGN noise dominates the performance and the results from previous work assuming infinite-bit ADCs \cite{Au-Yeung_TWC07, Mukkavilli_IT03} can apply.

In \figref{fig:MU_MISO_Rate_vs_SNR_Nt_4_K_2}, we show the achievable rates in a multi-user MISO channel. The number feedback bits is chosen as $B = 2(\Nt-1)b - 12 = 6b-12$. First, apart from the single-user case, there is a gap at high SNR between the case of perfect CSIT and limited feedback due to the inter-user interference. Second, the gaps between each pair of curves are all around $1.7$ bps/Hz, which verifies our analytical result in \eqref{eq:MU_MISO_scaling_law} stating that $B$ should increase $2(\Nt-1)$ bits if $b$ increases one bit. Third, at low SNR ($<0$ dB), the power loss is very small for three cases, which validates our results in \eqref{MU_MISO_Rate_loss_low_SNR} saying that the power loss is $10 \log_{10} \left(1 - 2^{-\frac{B}{\Nt-1}}\right)$ dB, which is around $-1.25$ dB when $B=6$, $-0.28$ dB when $B=12$, and $-0.07$ dB when $B=18$.

\section{Conclusions}

In this paper, we analyzed the achievable rate in MISO systems with finite-bit ADC and limited feedback. For both the single-user and multi-user channels, the results are similar to those with infinite-bit ADC at low SNR except for an additional power loss $10 \log_{10} \left(1-\eta_b\right)$ dB incurred by low resolution ADCs. At high SNR, however, the quantization noise dominates and therefore the results are very different from the case with infinite-bit ADCs. In the single-user channel, we found that the achievable rate saturates to a upper bound determined by signal-to-quantization ratio of the ADCs. In the multi-user case, we found that that the number of bits per feedback should increase linearly with the ADC resolution to limit the rate loss at high SNR.

\bibliographystyle{IEEEtran}
\bibliography{IEEEabrv,../../One_bit_quantization}
\end{document}